\documentclass{PoS}

\usepackage{caption}
\usepackage{subcaption}
\usepackage{graphicx}

\title{The optical noise monitoring systems of Lake Baikal environment for the Baikal-GVD telescope}

\ShortTitle{Lake Baikal optical noise}

\author{A.D.~Avrorin$^a$, A.V.~Avrorin$^a$, V.M.~Aynutdinov$^a$, R.~Bannash$^g$, I.A.~Belolaptikov$^b$, V.B.~Brudanin$^b$, N.M.Budnev$^c$, G.V.Domogatsky$^a$, A.A.Doroshenko$^a$, \speaker{R.Dvornick\'y}$^{,b,h}$, A.N.~Dyachok$^c$, Zh.-A.M.~Dzhilkibaev$^a$, L. Fajt$^{b,h,i}$, S.V.~Fialkovsky$^e$, A.R.~Gafarov$^c$, K.V.~Golubkov$^a$, N.S.~Gorshkov$^b$, T.I.~Gress$^c$, R.~Ivanov$^b$, K.G.~Kebkal$^g$, O.G.~Kebkal$^g$, E.V.~Khramov$^b$ , M.M.~Kolbin$^b$, S.O.~Koligaev$^j$,  K.V.~Konischev$^b$, A.V.~Korobchenko$^b$, A.P.~Koshechkin$^a$, A.V.~Kozhin$^d$, M.V.~ Kruglov$^b$, M.K.~Kryukov$^a$, V.F.~Kulepov$^e$, M.B.~Milenin$^a$, R.A.~Mirgazov$^c$, V.~Nazari$^b$, \fbox{A.I.~Panfilov$^a$}, D.P.~Petukhov$^a$ E.N.~Pliskovsky$^b$, M.I.~Rozanov$^f$, E.V.~Rjabov$^c$, V.D.~ Rushay$^b$, G.B.~Safronov$^b$, B.A.~Shaybonov$^b$, M.D.~Shelepov$^a$, F.~\u{S}imkovic$^{b,h,i}$, A.V.~Skurikhin$^d$, A.G.~Solovjev$^b$, M.N.~ Sorokovikov$^b$, I.~\u{S}tekl$^i$,  E.O.~Sushenok$^b$, O.V.~Suvorova$^a$, V.A.~Tabolenko$^c$, B.A.~Tarashansky$^c$, and S.A.~Yakovlev$^g$\\
$^a$ Institute for Nuclear Research, Russian Academy of Sciences, Moscow, 117312 Russia\\
$^b$ Joint Institute for Nuclear Research, Dubna, 141980 Russia\\
$^c$ Irkutsk State University, Irkutsk, 664003 Russia\\
$^d$ Institute of Nuclear Physics, Moscow State University, Moscow, 119991 Russia\\
$^e$ Nizhni Novgorod State Technical University, Nizhni Novgorod, 603950 Russia\\
$^f$ St. Petersburg State Marine Technical University, St. Petersburg, 190008 Russia\\
$^g$ EvoLogics Gmbh, Germany\\ 
$^h$ Comenius University, Mlynska Dolina F1, Bratislava, 842 48 Slovakia\\
$^i$ Czech Technical University in Prague, Prague, 128 00 Czech Republic\\
$^j$ Department of General and Applied Geophysics, Dubna State University, 141980 Russia\\
E-mail: \email{dvornicky@dnp.fmph.uniba.sk}
}

\abstract{
We present data on the luminescence of the Baikal water medium collected with
the Baikal-GVD neutrino telescope. This three-dimensional array of light sensors
allows the observation of time and spatial variations of the ambient light field.
We report on observation of an increase of luminescence activity in 2016 and 2018.
On the contrary, we observed  practically constant optical noise in 2017.
An agreement has been found between two independent optical noise data sets.
These are data collected with online monitoring system and the trigger system of the cluster.
}

\FullConference{36th International Cosmic Ray Conference -ICRC2019-\\
		July 24th - August 1st, 2019\\
		Madison, WI, U.S.A.}

\begin{document}

\section{Introduction}

Lake Baikal remains home to various unique species of plants and animals 
for millions of years, many of which are endemic. The vital conditions for
these creatures are horizontal and vertical water exchange processes, which supply
and distribute the oxygen and organic substances. The study of the hydrodynamic
processes in Lake Baikal are of particular interest for earth and life sciences.
Beyond limnology, it may improve our understanding of the hydrodynamics of seas 
and oceans. 
    
The next generation neutrino telescope Baikal-GVD is placed in the southern basin 
of Lake Baikal about 3.6\,km from shore at a depth of 1\,366\,m.  The main goal 
of the experiment is the detection of high energy astrophysical neutrinos, 
whose sources remain still unknown. In particular, the aim is the registration 
of the Cherenkov radiation emitted when secondary charged particles, created 
in the reactions of neutrinos with surrounding medium, are passing through 
the deep water in Lake Baikal. The detector itself is a three-dimensional array 
of photo-sensitive components called optical modules (OMs). 
A fully independent unit called cluster consists of 288 OMs attached on 8 strings, 
7 peripheral strings surrounding the central one with a radius of 60\,m.
Each string carries 36 OMs with 15\,m vertical spacing. The top and the bottom OMs 
are located at depths of 750\,m and 1\,275\,m, respectively. In 2016, the first cluster
"Dubna" has been deployed. In the two subsequent winter expeditions of 2017 and 2018, 
two more clusters have been deployed. Another two clusters have been deployed 
during the winter expedition of 2019. In recent, the total number of the deployed 
clusters is five \cite{icrc19simkovic}.

Apart of Cherenkov radiation, also the ambient background light is registered. 
The amount of the registered background light is derived from the photo-multiplier
noise rates from each particular OM. There are two independent ways of collecting
the data. The trigger system of every cluster is designed in such a way 
that signals from each OM in a time window of 5 $\mu s$ are stored, if a trigger 
condition is fulfilled \cite{icrc19khramov}. In this way, we obtain the data on count rates 
of pulses registered by OMs. Besides this way, there is an online monitoring system,
which collects data from the OM controller electronics placed inside every OM.  
The origin of the background noise rates is mainly associated with the luminescence of the Baikal 
water. By means of the two independent systems, the light registration is almost continuous. 
In this article, we present some selected results on luminescence in Lake Baikal.

\section{Optical activity of the Baikal water}

\begin{figure}[ht]
\begin{center}$
\begin{array}{cc}
\includegraphics[width=74mm]{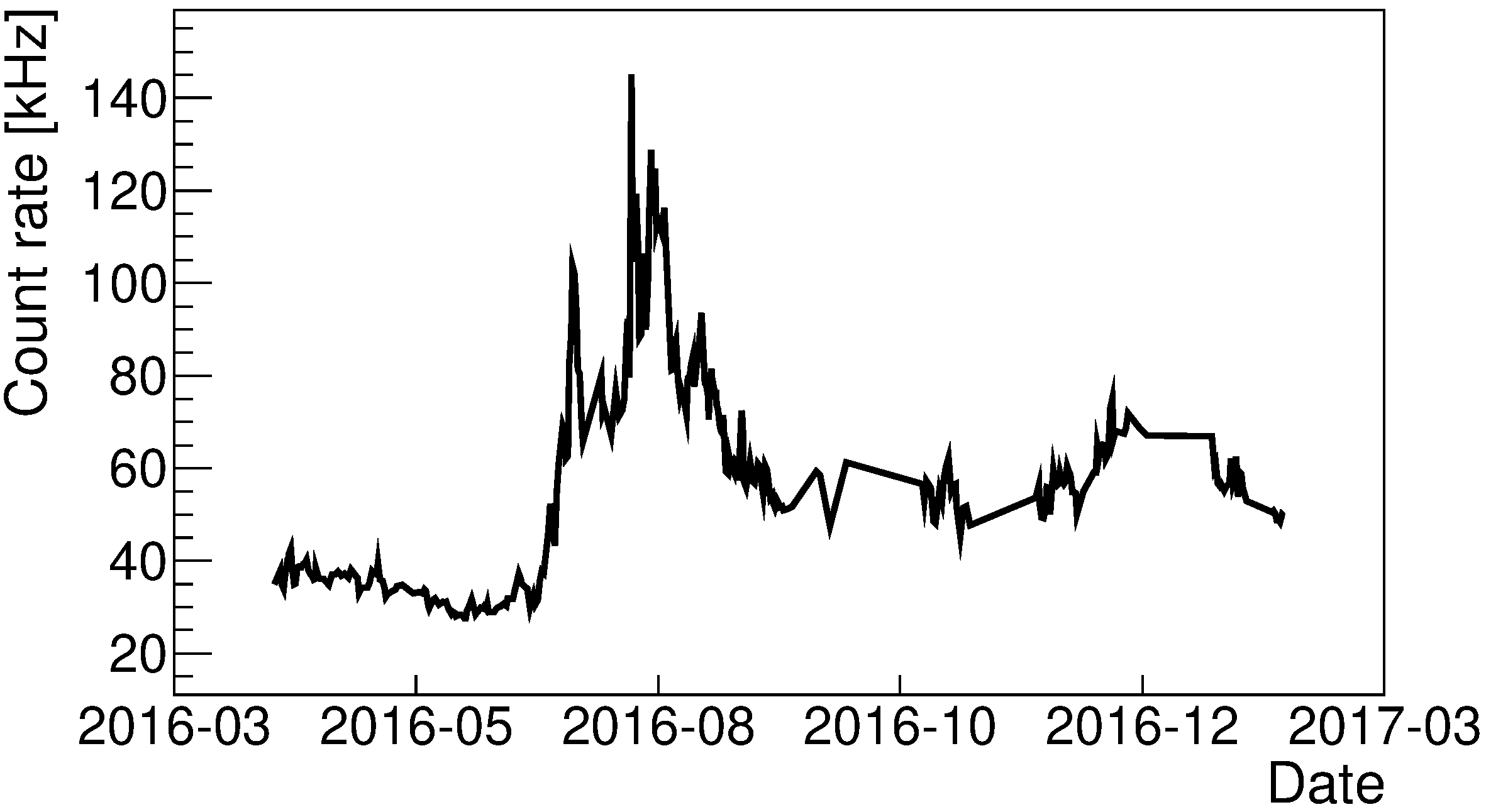}  &
\includegraphics[width=74mm]{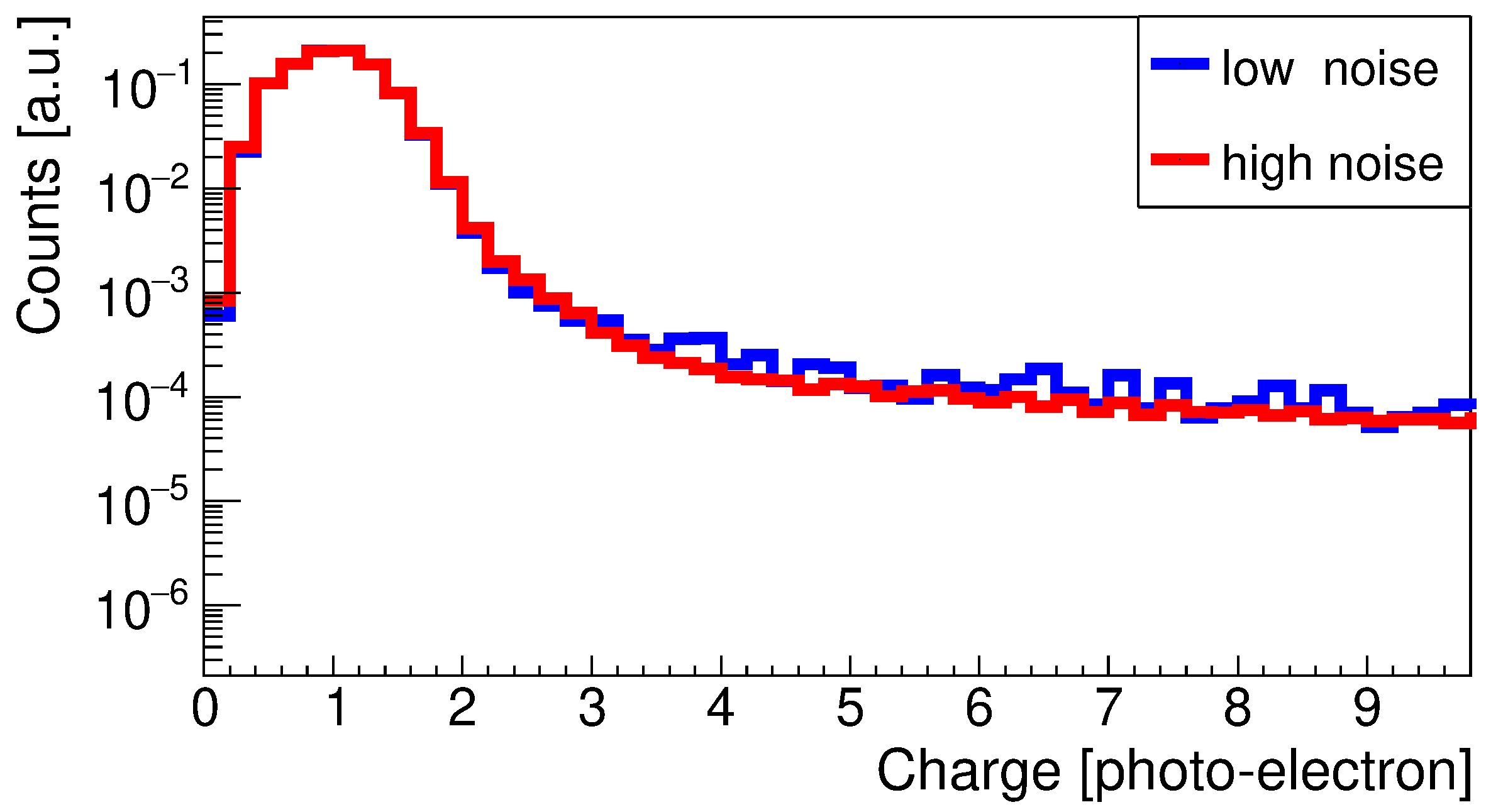}
\end{array}$
\end{center}
\caption{Left panel: Count rate evolution  of a selected OM from April 2016 till February 2017.
Right panel: Charge distribution of registered pulses in units of photo-electrons.
\label{fig.1}}
\end{figure}

Baikal-GVD is designed to detect the Cherenkov light from charged particles.
In open water, light not related to relativistic particles constitutes an unavoidable 
background to the Cherenkov light. Therefore studies of the related light fields 
are of crucial importance. The photon flux from the sunlight below a depth of $\sim$ 700 m 
is negligible as shown in previous work \cite{luminescence98}. 

\subsection{Background light - features}

In Fig.\ref{fig.1} (left panel), we present data on count rates for a selected OM 
for April 2016 -- February 2017. There are two periods of relatively stable
optical background noise, which are intermitted by increased optical activity.
The charge distribution of the noise pulses is displayed in Fig.\ref{fig.1} (right panel).
We stress that the charge distribution remains unchanged in different periods of the
optical activity. Our measurements are performed with a threshold of half a single 
photo-electron charge. In this way, the dark noise of the photo-multiplier is significantly 
suppressed. We note that by setting the threshold to one photo-electron the background 
count rate is reduced by a factor of two. The one photo-electron background is well
correlated with the half photo-electron background. The count rates in both cases exhibit
the same modulation of the relative amplitude. We clearly see that the major contribution 
comes from single photo-electron pulses. 

\subsection{Background light - time variations}

\begin{figure}[h]
\begin{center}$
\begin{array}{cc}
\includegraphics[width=74mm]{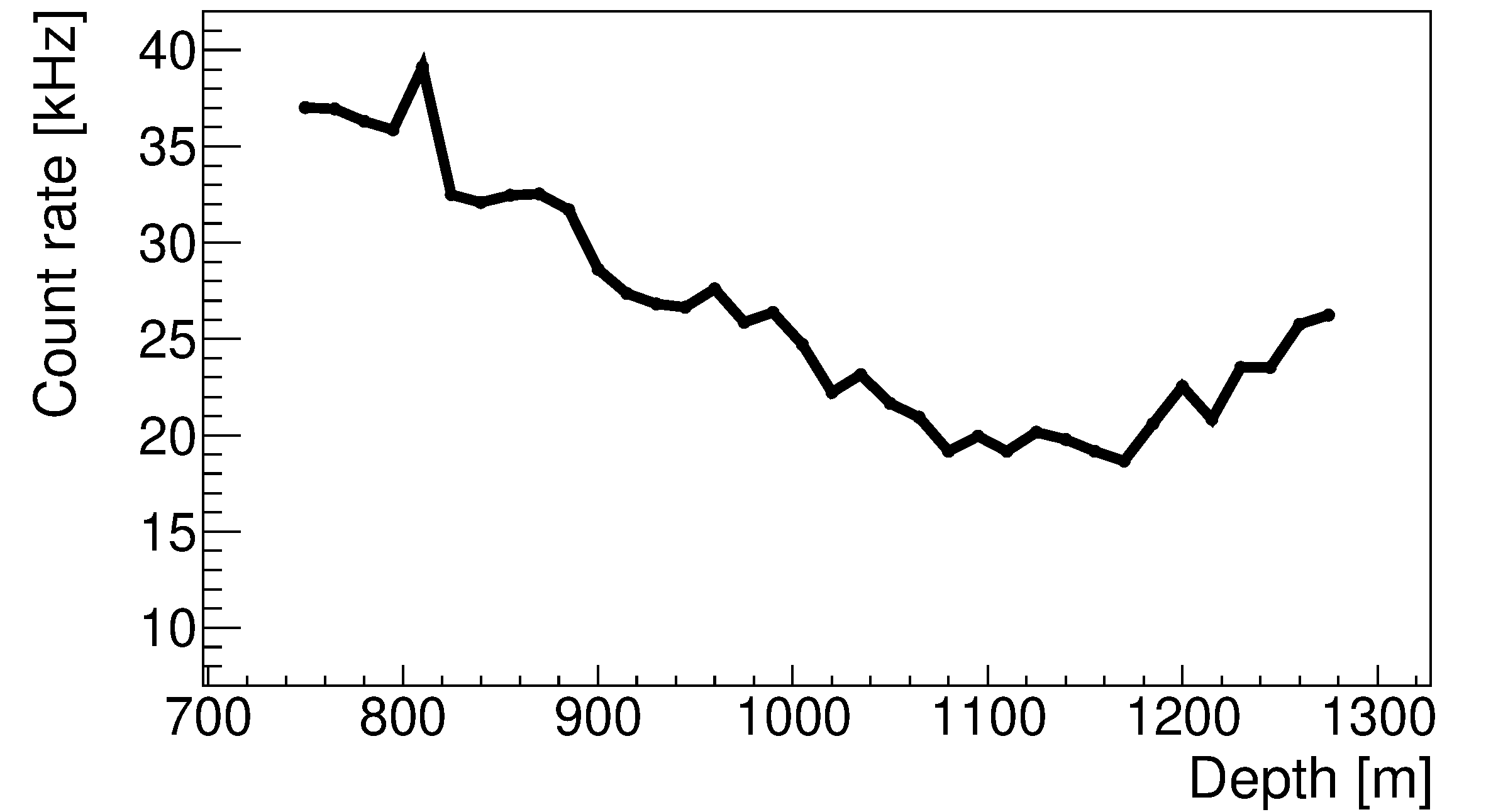}   &
\includegraphics[width=74mm]{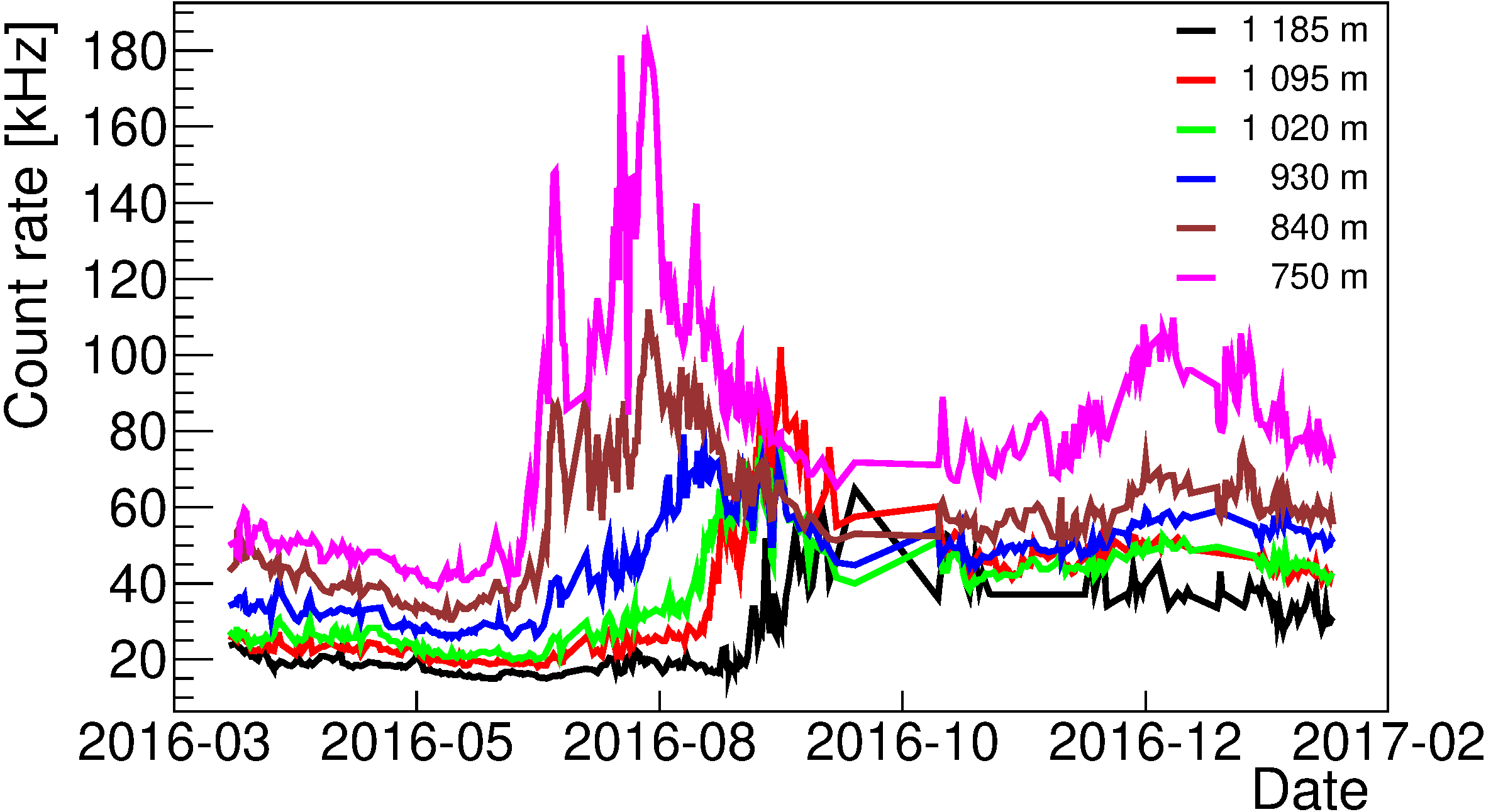}
\end{array}$
\end{center}
\caption{Left panel: Count rates averaged for each depth
over the OMs from different strings as a function of depth.
The lake bed is at $1\,366$\,m depth. Data from June 2016. 
Right panel: Count rates for the OMs of the same string at different depths. 
For the sake of simplicity, we show only six of 36 OMs, placed 
at depths of 750, 840, 930, 1\,020, 1\,095, and 1\,185 meters.
\label{fig.2}}
\end{figure}

The depth dependence of the ambient light field is the same for all eight strings of a cluster.
By averaging the count rates over the OMs at the same horizon, we obtain the depth dependence
of the background light noise. The average count rates versus the depth are presented 
in Fig.\ref{fig.2} (left panel). The analysed data are from June of 2016. This is the period of 
the lowest optical activity. We note that the pattern remains the same for other periods of
stable noise activity.

During the period of increased activity, the depth dependence is displayed in Fig. \ref{fig.2} (right panel). 
The appearance of the outbreak maximum depends on time, starting with the top modules. 
Indeed, we observe a layer of highly luminescent water moving from the top to the bottom 
of the lake. By comparing the maximum for different depths, we obtain a velocity profile 
of the flows. In the beginning of August, the estimated speed reached its maximal value of 
$\sim$45\,m/day, while it remained almost constant ($\sim$8\,m/day) till the end of September,
i.e. when the activity asymptotically reached the background plateau. The observed pattern is 
similar to previous investigations with NT200 detector (see \cite{nonstationarity}).

\begin{figure}[h]
\begin{center}$
\begin{array}{cc}
\includegraphics[width=74mm]{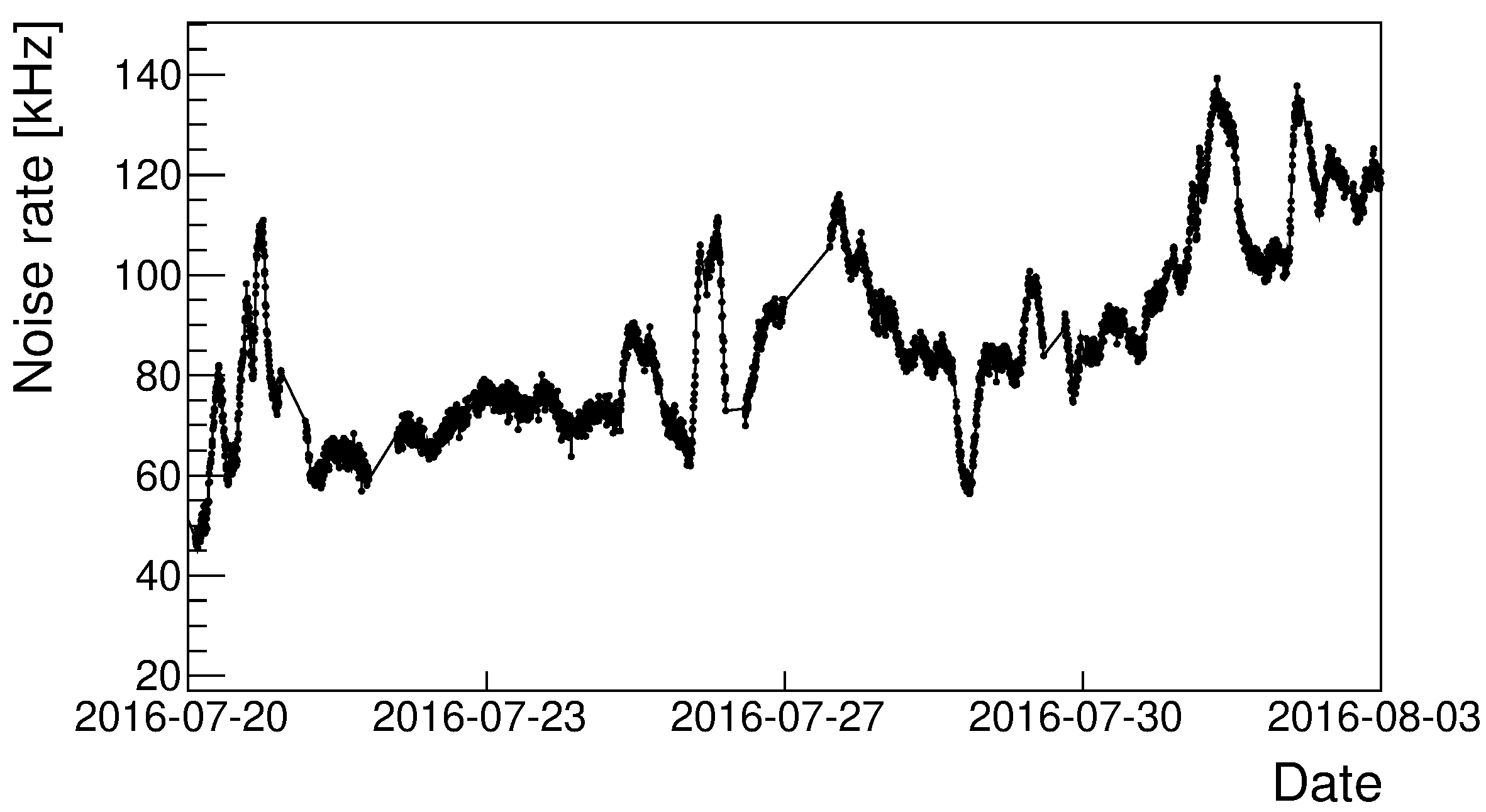}    &
\includegraphics[width=74mm]{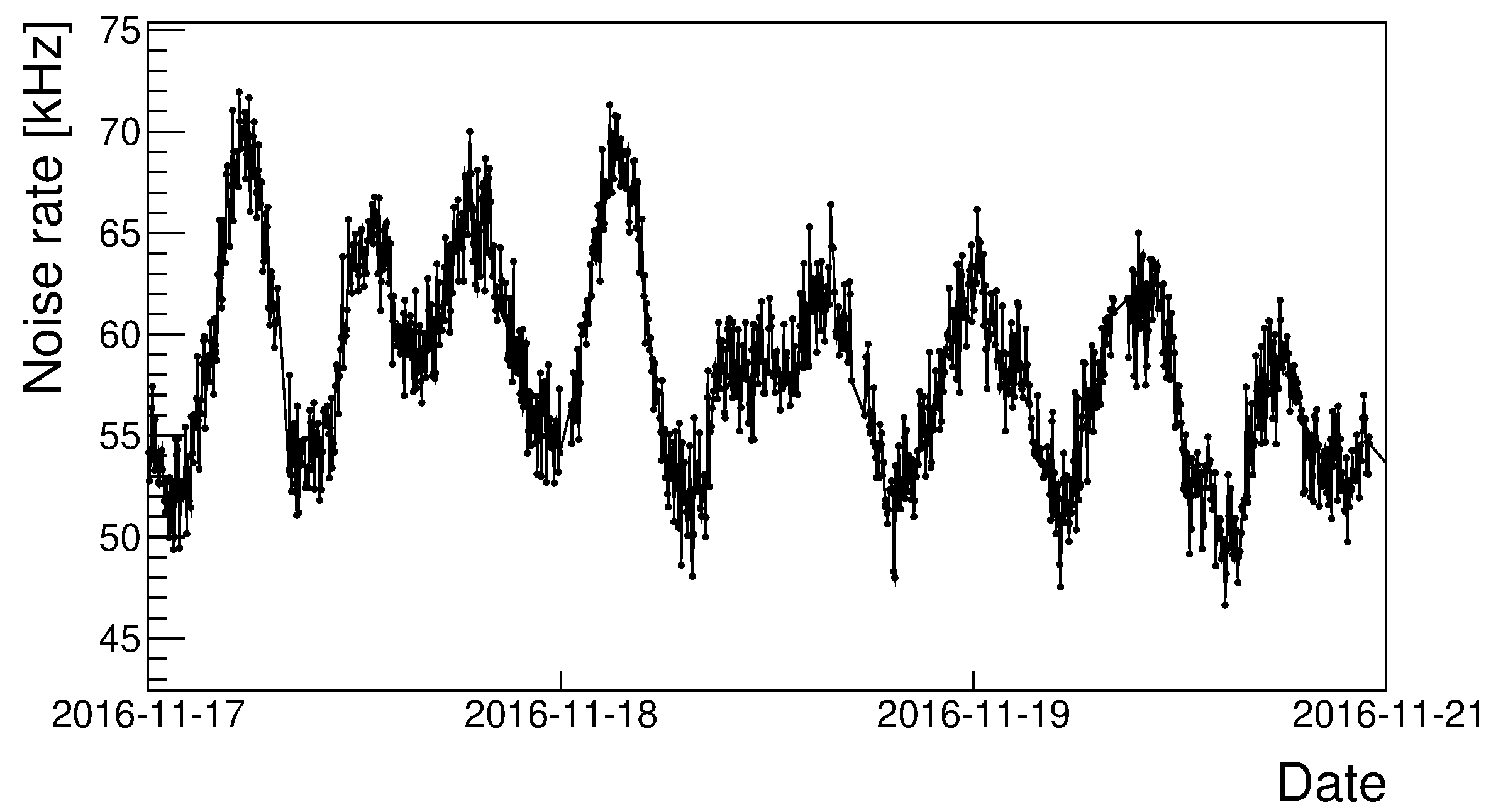}
\end{array}$
\end{center}
\caption{ Left panel: Count rates for a particular OM during the optically highly active period,
namely from July till September 2016. A sudden outburst of the count rates is noticeable.  
Right panel: An example of the regular modulation of noise rates. 
Data are from the period of stable plateau, namely from October 2016 till February 2017. 
\label{fig.3}}
\end{figure}

\begin{figure}[h]
\begin{center}$
\begin{array}{cc}
\includegraphics[width=74mm]{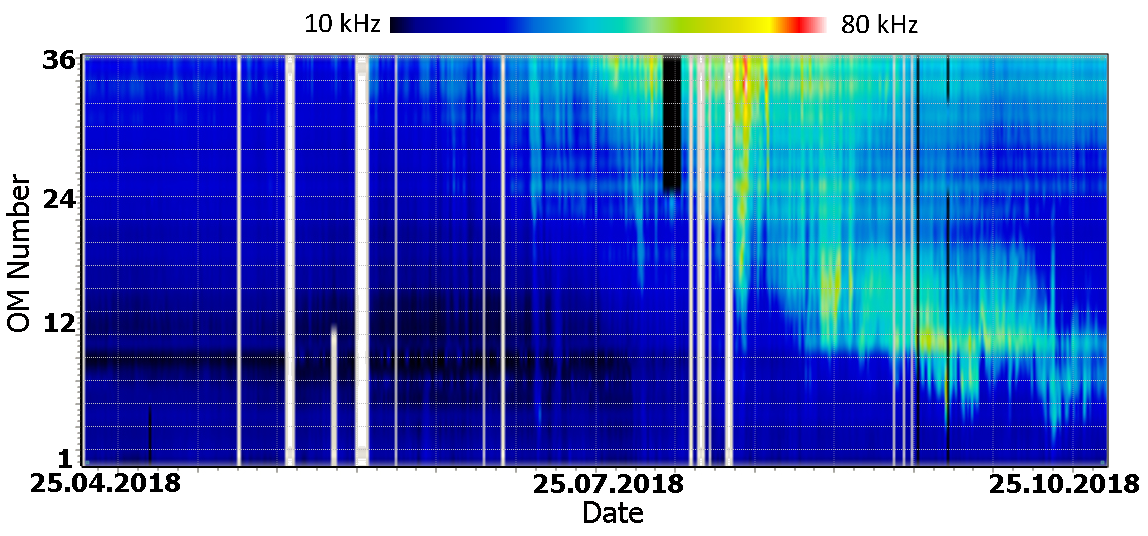}     &
\includegraphics[width=74mm]{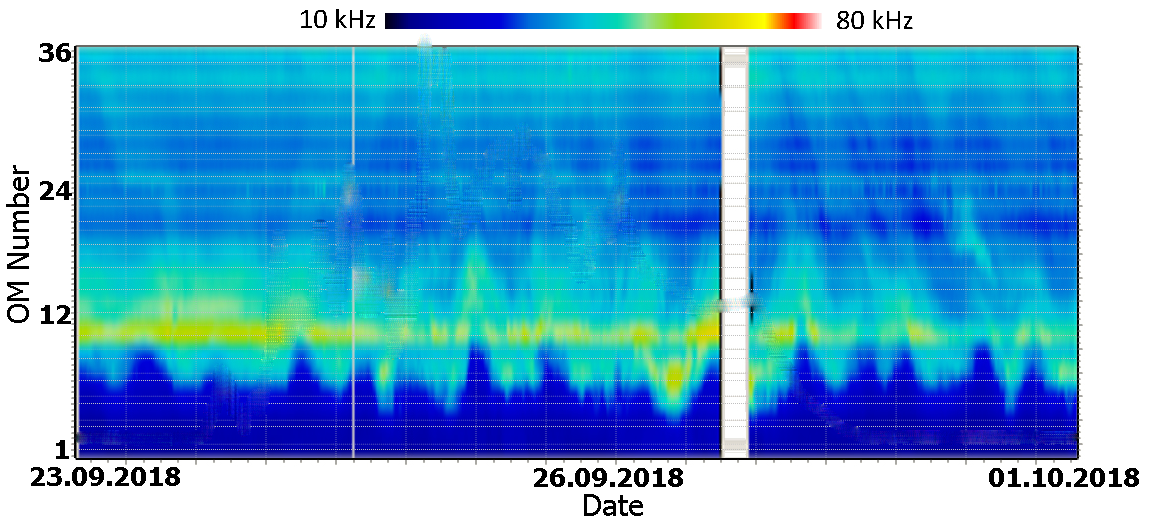}
\end{array}$
\end{center}
\caption{Left panel: Count rates for 36 OMs at the same string. The bottom and the top
OM are labeled as No.1 and No.36, respectively. Collected data are from the year 2018. 
High (Low) noise rates are presented in red (blue).
Right panel: The same as in left panel for a particularly selected time window,
when the effect of regular modulation is clearly manifested.  
\label{fig.4}}
\end{figure}

The time evolution of the count rates, as shown in Fig. \ref{fig.2} (right panel),
exhibits sharp changes on top of relatively continuous smooth optical background.
The effect is more visible in particularly selected time window displayed in Fig. \ref{fig.3} (left panel).
The amplitude of these sudden changes reached almost 50\,kHz. The duration of such variations
which distort the smooth background ranges typically from several hours up to a few days. 
We note that effect is present in July -- September 2016, i.e. the period of increased luminescent 
activity. However, the period of relatively stable plateau (October 2016 -- February 2017)
shows (see Fig. \ref{fig.3}, right panel) regular modulation of noise rates. The period of these
modulations is quite stable and varies from 10-12 hours. We stress that these waves are 
probably the manifestation of the internal waves in the lake. The end of these modulations
cannot be determined as far as the measurement during the year is interupted by the winter 
expedition (For further details see \cite{icrc19simkovic}). On the other hand, 
we observed a practically constant background noise without a period of high 
luminescence activity in 2017. 

However, the noise rates in 2018 exhibit similar pattern to that already described above for
the year 2016. In Fig. \ref{fig.4} (left panel) we evidently see a luminescent layer moving 
from the top to the bottom of the lake. We observed the regular modulation of the noise rates
again, as shown in Fig. \ref{fig.4} (right panel). Firstly, the modulations appeared on top OMs
in June 2018 and persisted till the end of October 2018. The maximal amplitude reached 70\,kHz.

\section{Torrent currents in Lake Baikal}

\begin{figure}[!h]
\begin{center}$
\begin{array}{cc}
\includegraphics[width=73mm,height=40mm]{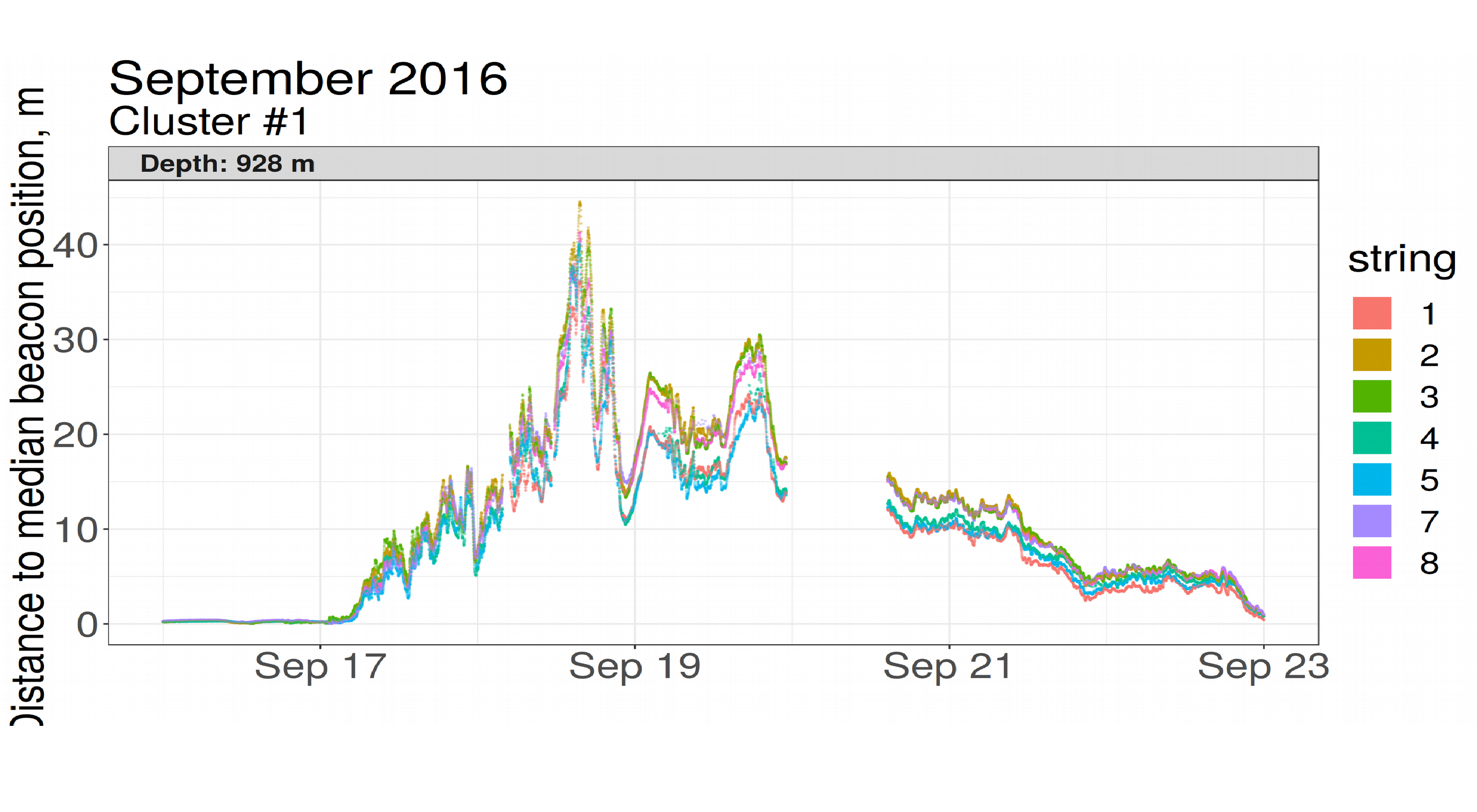}    &
\includegraphics[width=73mm,height=40mm]{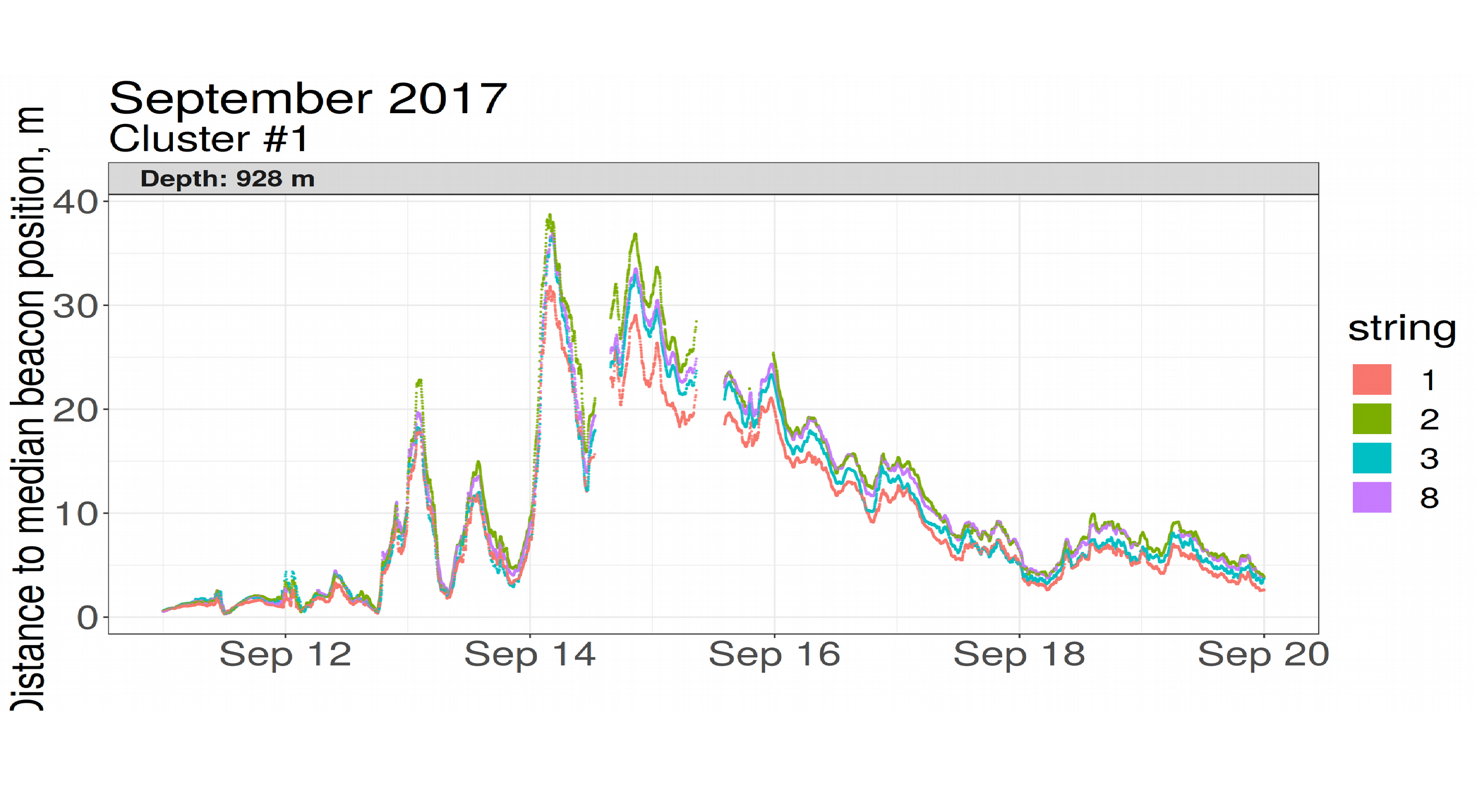}  \\
\includegraphics[width=71mm]{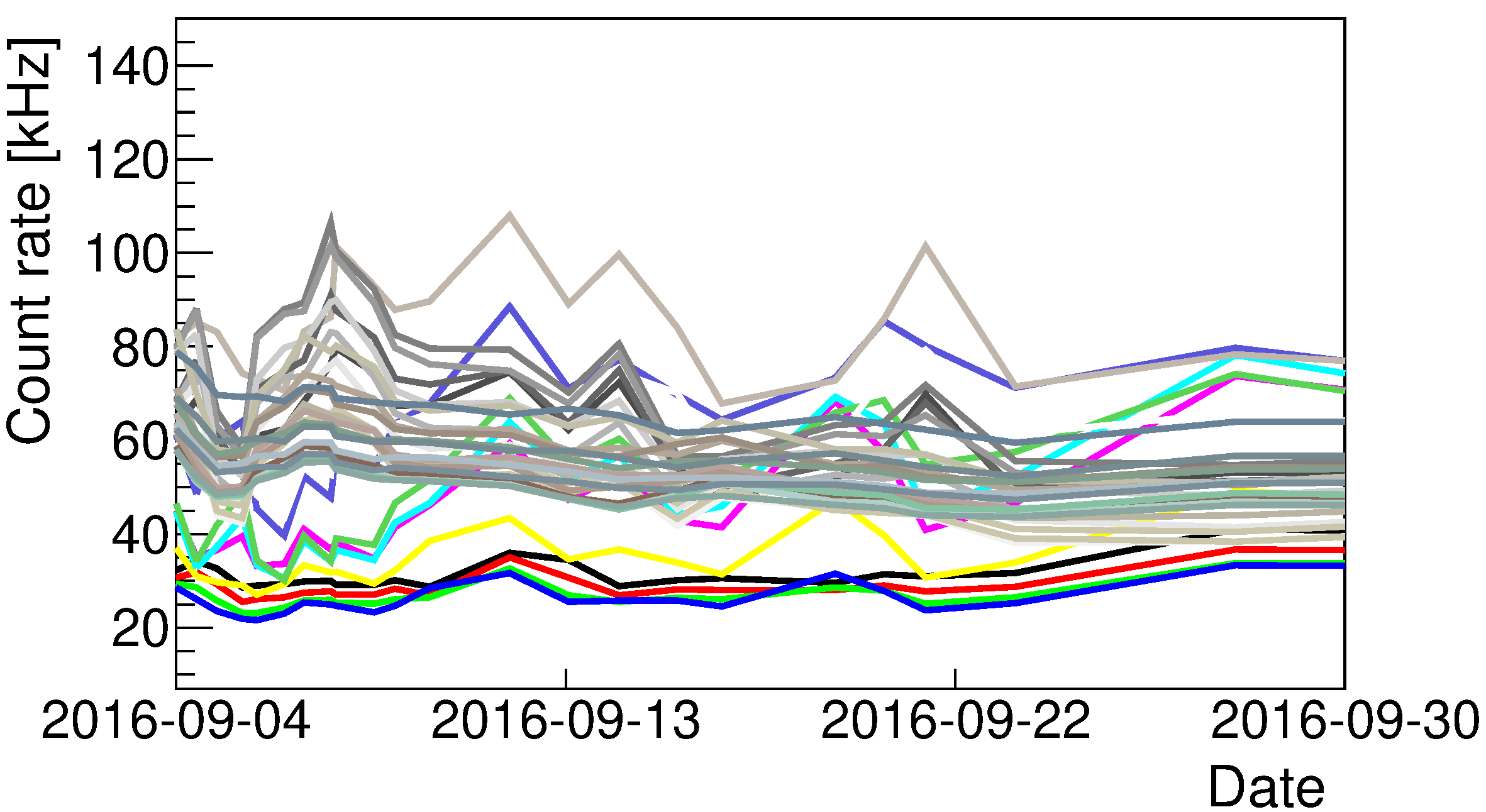}     &
\includegraphics[width=71mm]{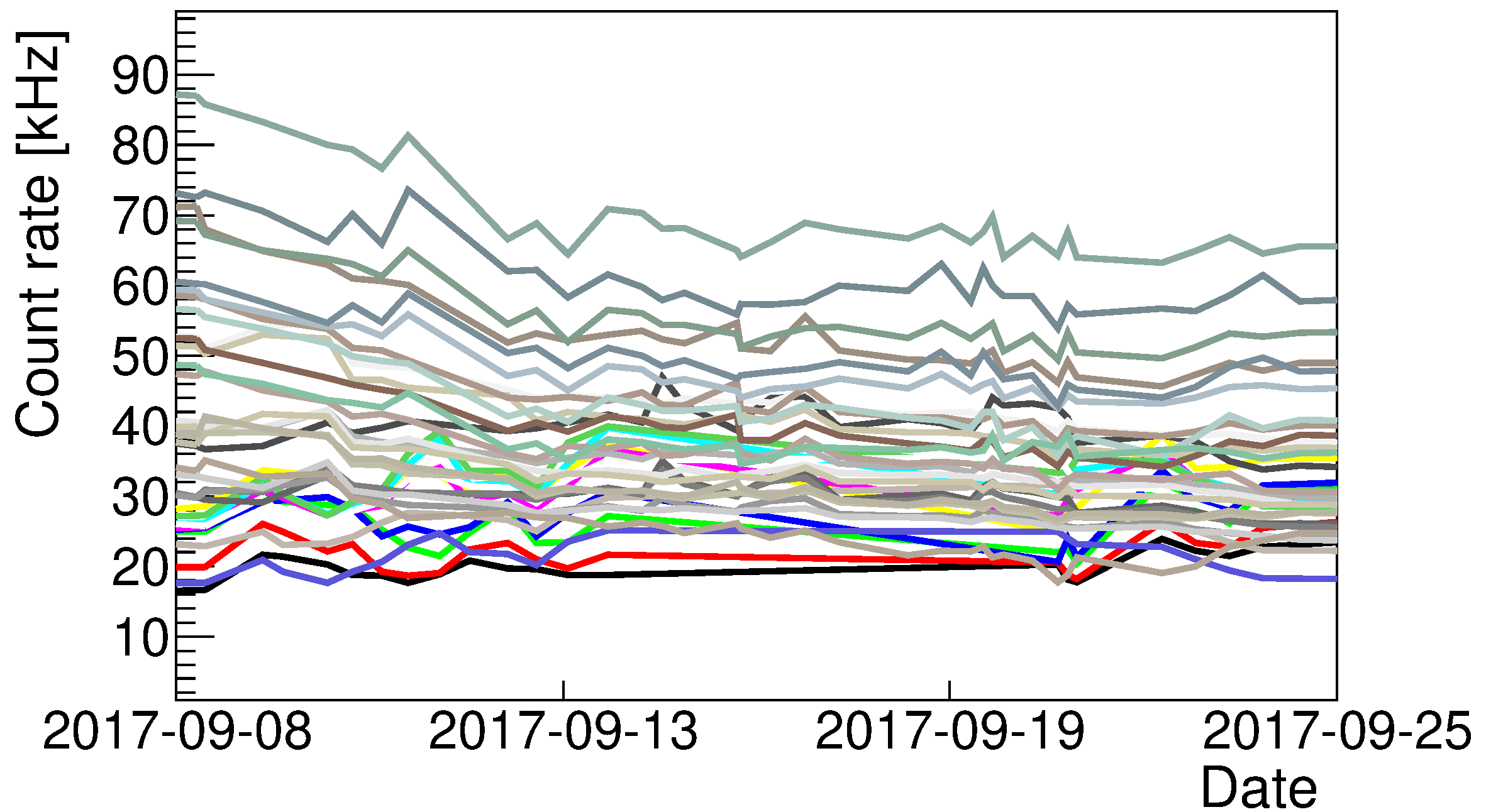}   
\end{array}$
\end{center}
\caption{ Deviations of beacons from their median positions at different strings 
for data of autumn 2016 (left upper panel) and autumn 2017 (right upper panel).
In these periods deviations of strings from their median positions were extremal.
The count rates of 36 OMs at the same string for data taken in the period when deviations
of the string from the median position were extremal (left lower panel) and for data
taken in 2017 (right lower panel). 
\label{fig.5}}
\end{figure}

Due to the currents in Lake Baikal, the string geometry deviates from its
vertical direction. To take these deviations into account, an acoustic positioning system
for Baikal-GVD has been developed (For more details see the contribution to this conference
\cite{acou}). Our observations show two periods of extreme deviations of the strings, 
in September of 2016 and of 2017. Torrent flows in the lake may produce a remarkable tilt
of the string from its vertical position, two examples of which are displayed in 
Fig. \ref{fig.5} (left and right upper panels). For the same period, we present the data
on count rates in Fig. \ref{fig.5} (left and right lower panels). We do not find a correlation 
between the torrent flows of the deep water and the luminescence activity of the lake.

\section{Noise monitoring system}

\begin{figure}[h]
\begin{center}
\includegraphics[width=120mm,height=70mm]{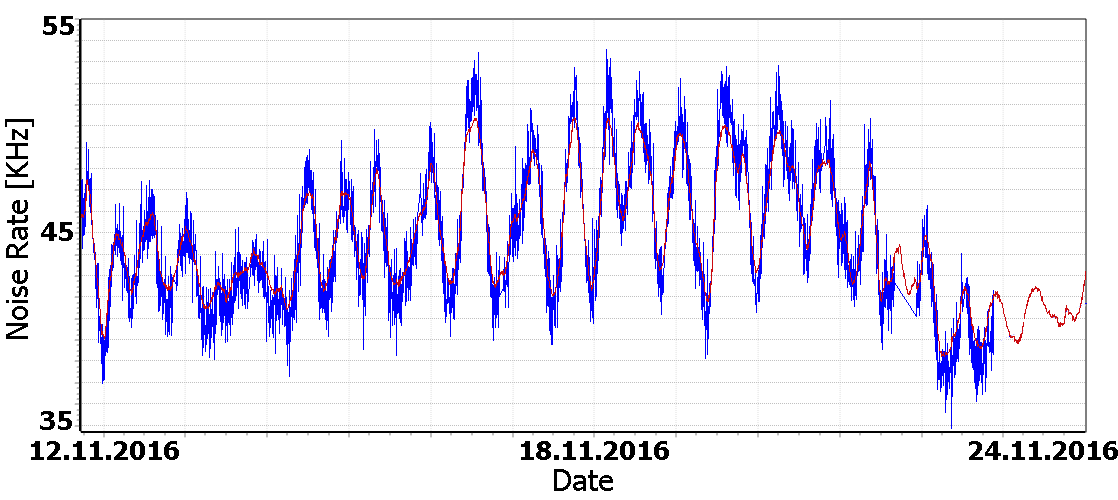} 
\end{center}
\caption{Noise rates acquired with online monitoring system and the cluster trigger system.
Data are properly scaled as in each of systems different pulse thresholds are considered. 
A rather good agreement between the two datasets is remarkable.
\label{fig.6}}
\end{figure}

The online monitoring system of the telescope is designed for the continuos registration
of the measurement conditions for each OM. For more details see \cite{icrc17golubkov}.
Here, we put emphasis on the photo-multiplier noise rate only, which is the subject
of our interest. A counter of nanosecond pulses is built in the OM electronics. 
In this way, the count rates are interrogated in regular time windows almost continuosly.
It is worth to mention that the noise rates obtained from monitoring system are, 
unlike the data acquired from the trigger system of the cluster, completely independent
of the trigger system settings and parameters. However, the count rates data are collected
with different thresholds of registered pulses in these two ways. Therefore, one needs
to rescale the noise rates when comparing the data obtained from monitoring system versus
the data acquired by the trigger system of the cluster. In Fig.(\ref{fig.6}), 
we present a comparison between the noise rates obtained in these two different ways. 
We see that both data sets agree well with each other.

\section{Conclusions}

We have presented data on the luminescence in Lake Baikal which have been  
collected by the Baikal-GVD neutrino telescope. We found an increase of the luminescence
activity intermitting periods of relatively stable optical background in 2016 and 2018. 
On the contrary, we observed practically constant background noise without a period of
high luminescence activity in 2017. Moreover, we find that the maximum of the optical 
activity observed in 2016 propagated from top to bottom, with a maximum speed of 45\,m/day. 
We did not find a correlation between the torrent flows and the increase of the luminescence activity.
An agreement between the noise rates datasets obtained from cluster trigger system and 
the online monitoring system has been found.

\section{Acknowledgements}

This work was supported by the Russian Foundation for Basic Research (Grants 16-29-13032, 17-0201237).

\end{document}